\pdfoutput=1 
\documentclass[prb,aps,twocolumn,10pt,letterpaper,citeautoscript,
			    floatfix,superscriptaddress,raggedbottom]{revtex4-1}
\usepackage{graphicx}
\usepackage{amsmath}
\usepackage{amssymb}
\usepackage[bookmarks=false,%
			pdfstartview=FitH]{hyperref}
\usepackage[protrusion=true,expansion]{microtype}

\begin{document}
\title{%
 	Octahedral rotation-induced ferroelectricity in cation ordered perovskites}
\author{James M.\ Rondinelli}
	\email{jrondinelli@coe.drexel.edu}
	\affiliation{Department of Materials Science \& Engineering,\!
	Drexel University,\! Philadelphia,\! PA 19104,\! USA}%
\author{Craig J.\ Fennie}
	\email{fennie@cornell.edu}
	\affiliation{School of Applied \& Engineering Physics, Cornell University, 
	Ithaca, New York 14853, USA}
\date{\today}


 
%
\maketitle

\begin{figure}
\centering\vspace{3pt}
\includegraphics[width=0.44\textwidth]{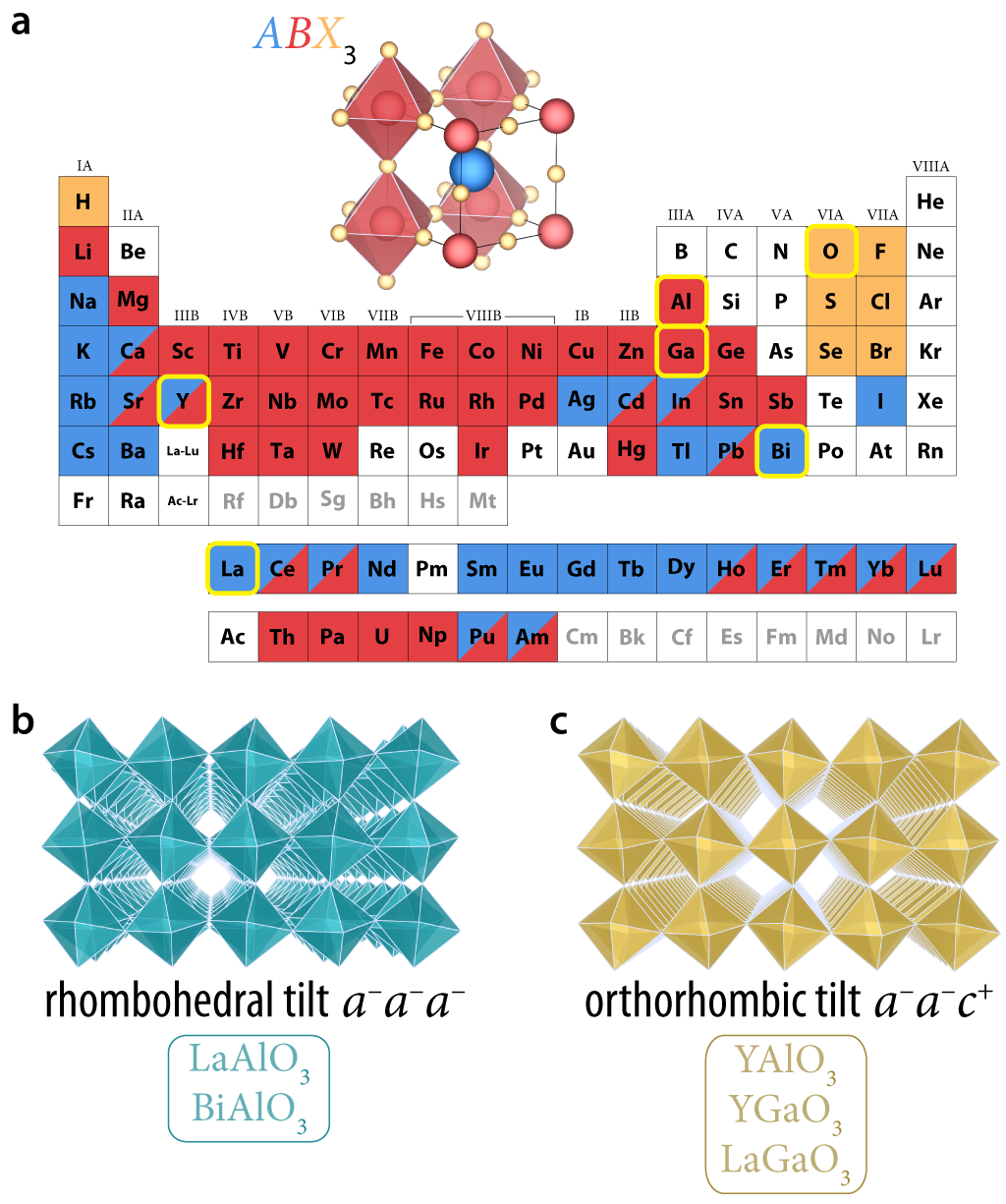}\vspace{-8pt}
\caption{\label{fig:perovskite}The ubiquitous octahedral rotations 
available to perovskites regardless of chemistry preserve inversion symmetry. 
(a) The adaptable $ABX_3$ perovskite structure permits a vast number 
of chemical compositions, indicated by the color-coded 
periodic table, and therefore also exhibits nearly every conceivable 
materials property. 
The \emph{oxide} material class, especially the compositions studied here (listed below), 
exhibits a network of corner-connected 
$B$O$_6$ octahedra which are often distorted from the cubic 
structure shown in (a).
The pervasive low symmetry (centrosymmetric)  rhombohedral (b) and orthorhombic (c) 
structures that result are due to rotations of the  $B$O$_6$ building blocks.
Following Glazer notation\cite{Glazer:1972}, the $a^-a^-a^-$  octahedral tilt pattern in (b) consists of rotations of adjacent $B$O$_6$ octahedra that are out-of-phase (--) and of equal magnitude along all Cartersian directions.
The $a^-a^-c^+$ tilt pattern (c) is similar as it possesses  
out-of-phase rotations about two directions with equal magnitude (perpendicular 
to the projection illustrated), but exhibits in-phase rotations (+) about 
the remaining Cartesian direction with different amplitude (page normal).
Periodic table adapted after Ref.~\onlinecite{Schlom/Hawley:1998}.}
\end{figure}

Increasing demands for electric field-tunable electric, magnetic, and orbital (EMO) materials has renewed interests in ferroelectricity and its coupling to EMO properties. 
Materials where a spontaneous electric polarization $P$, which naturally couples to and is switched by an electric field, strongly interacts with EMO degrees of freedom provide a platform to realize novel electric-field-controllable,  low-power multifunctional devices such as ultra-fast Mott-based devices.\cite{Newns/Schrott_et_al:1998,Auciello/Scott/Ramesh:1998,Ahn/Triscone/Mannhart:2003,Rini/Cavalleri_et_al:2007,Scott:Book,Hormoz/Ramanathan:2010,Takagi/Hwang:2010,Mannhart/Schlom:2010}
In the versatile class of complex $AB$O$_3$ perovskite oxides (Fig.~\ref{fig:perovskite}), 
ferroelectric polarizations are usually induced by polar displacements of 
second-order Jahn-Teller (SOJT) active cations\cite{Burdett:1981,Kunz/Brown:1995,Bersuker:2001} on the $B$-site (commonly $d^0$ transition metals ions) and/or on the $A$-site (lone-pair active cations).
The cation displacements that induce ferroelectricity, however, are often incompatible with and/or weakly coupled\cite{Thomas:1996b,Ghita/Singh_etal:2005} to EMO-derived material properties such as  electronic bandwidths,\cite{Torrance/Niedermayer_etal:1992,Eng/Woodward:2003} magnetic interactions\cite{Subramanian_et_al:1999,Zhou/Goodenough:2006} and critical transition temperatures.\cite{Millis:1998,Goto/Tokura_et_al:2004} 
Rotations of corner-sharing $B$O$_6$  octahedra, however, directly alter these macroscopic 
properties,\cite{Imada/Fujimori/Tokura:1998} because they buckle the inter-octahedral $B$--O--$B$ bond angles which mediate the interplay of EMO  degrees of freedom.
Thus in perovskites, although the vast number of chemical compositions available (Fig.~\ref{fig:perovskite}a) facilitates nearly every conceivable material property, the local polar cation displacements necessary for ferroelectricity often occur independently of the property-controlling and pervasive\cite{Glazer:1972,Woodward:1997a,Woodward:1997b} $B$O$_6$ rotations in the extended framework---rotations account for nearly 75\% of all rhombohedral, Fig.~\ref{fig:perovskite}b, and orthorhombic, Fig.~\ref{fig:perovskite}c, structures.

\begin{figure*}
\includegraphics[width=0.75\textwidth]{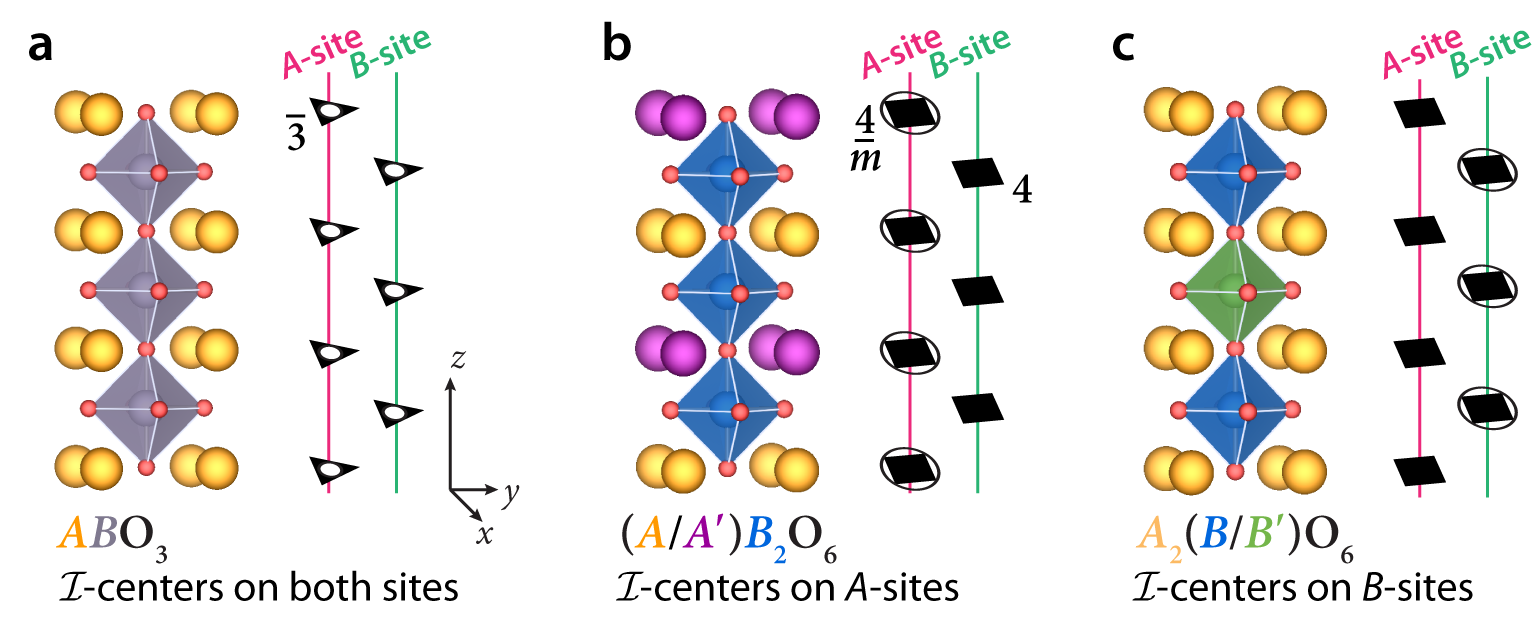}\vspace{-7pt} 
\caption{\label{fig:ingredients} Illustration of the Chemical Criterion for 
rotation-induced ferroelectricity in layered perovskite superlattices constructured from two different $AB$O$_3$ perovskite materials.
In (a) bulk $AB$O$_3$ perovskites inversion ($\mathcal{I}$) centers are found on both the $A$- and $B$-sites; the highest site-symmetry operator being a three-fold rotoinversion ($\bar{3})$.
Cation ordering in layered perovskites, however, lifts the inversion centers on
the $B$-site (leading to a 4-fold rotation) in the $A/A^\prime$ layered perovskites (b) and on the $A$-site in the $B/B^\prime$  (c) structures.
Inversion  remains through the $\frac{4}{m}$ operation found on the remaining $A$-site  and $B$-site, respectively. 
Since rotations of octahedra preserve the inversion on $B$-sites yet can remove it on the $A$-sites, only $A/A^\prime$ 
support this form of hybrid improper ferroelectricity.
}
\end{figure*}

The strong coupling of oxygen octahedral rotations to the EMO properties in the perovskite structure presents an opportunity to create novel multifunctional materials that respond to external electric fields. 
The steric linkage of the  $B$O$_6$ framework, however, constrains rotations of octahedra in the same plane to have equal magnitudes but opposite rotational ``sense,'' thereby preserving inversion symmetry\cite{Stokes/Kisi_et_al:2002}. Thus collective rotation patterns do not induce an electric polarization nor directly couple to an electric field.
To circumvent this reality, synthetic chemical routes have focused on turning the \textit{centric} $B$O$_6$ octahedra into \textit{acentric} structural units by partial halide substitution.\cite{Maggard/Poeppelmeier:2001,Welk/Poeppelmeier:2002,Marvel/Poeppelmeier:2007} Progress in \textit{cis}- and \textit{trans}-ordering of the halide anions throughout all acentric units, however, remains slow and ferroelectric switching even more challenging.\cite{Heier/Poeppelmeier:1998,Maggard/Poeppelmeier:2003,Yang/Attfield:2011}
Recently, a new strategy involving the layering of perovskites  blocks\cite{Bousquet/Ghosez_et_al:2008,Etxebarria/Perez-Mato:2010,Benedek/Fennie:2011,Fukushima/Picozzi:2011}  has lead to the realization of ferroelectric systems whose electric polarizations can be completely accounted for by centric  rotations.  The mechanism has been referred to as hybrid improper  ferroelectricity, however, no comprehensive  design prescription applicable to multiple  chemistries has been proposed.
In this work, we present a materials design strategy to realize ferroelectricity in layered perovskite superlattices constructed from two different $AB$O$_3$ materials, neither of which are ferroelectric in bulk.
The ferroelectricity is of a hybrid improper form~\cite{Bousquet/Ghosez_et_al:2008,Etxebarria/Perez-Mato:2010,Benedek/Fennie:2011}, where the polarization $P$ can be induced by the coexistence of \emph{two} octahedral rotation patterns of different symmetry.
The design rules are constructed from group-theoretical methods combined with principles of perovskite crystal chemistry and are therefore completely general. In particular for  ($AB$O$_3$)$_1$/($A^\prime B^\prime$O$_3$)$_1$  superlattices, a layered $A/A^\prime$ cation ordering, designated as a  \emph{chemical criterion}, is required to realize this new form of ferroelectricity, whereas optimal materials selection, corresponding to an \emph{energetic criterion},  is based solely on the rotation patterns present in the parent single phase $AB$O$_3$ and $A^\prime B$O$_3$ bulk perovskites.  We find that the parent materials do not need to exhibit ferroelectricity, but rather that each constituent should have a tendency to distort in the orthorhombic $Pnma$ ($a^-a^-c^+$ in Glazer notation, see  Fig.~\ref{fig:perovskite}c) octahedral tilt pattern in bulk.
Using \textit{ab initio} first-principle techniques we test the design rules against artificially layered gallate and aluminate perovskites. (Note that our focus on $d^0$ cations is to avoid the known challenges of first-principles theory in treating materials with partially filled $d$- and/or $f$-shells and is not a limitation of our strategy.)
These materials design rules  are capable of rapidly guiding the discovery of numerous unknown multifunctional materials as  the $A/A^\prime$ cation layered ordering is highly amenable to advanced solid-state synthesis\cite{Graham/Woodward:2010,Dachraoui/Greenblatt:2011} and layer-by-layer\cite{Zubko/Triscone_et_al:2011} deposition techniques. Additionally, the targeted $a^-a^-c^+$ octahedral tilt pattern is the most common type observed  in perovskites.\\

\noindent %
{\sffamily \bfseries Structure--property guidelines$\quad$}\\ \noindent
Hybrid improper ferroelectricity (HIF) has been shown to arise from a 
peculiar trilinear lattice coupling 
term\cite{Bousquet/Ghosez_et_al:2008,Benedek/Fennie:2011}
in the thermodynamic free energy, $g P (Q_1 Q_2)$,  where the 
polarization $P$ is induced by the product of two translation symmetry-breaking 
lattice modes, $Q_1$ and $Q_2$, of different symmetry. The interaction strength of the coupled modes is described by the temperature-independent coefficient $g$.
The principal design challenge is identifying the universal 
structural-chemical requirements within a class of materials that 
allows the non-polar lattice modes to induce ferroelectricity.
Here, we focus on octahedral rotation-induced ferroelectricity
in perovskite-structured oxides due to the abundance of 
$Q$ modes describing the $B$O$_6$ connectivity.
Our crystal--chemistry design approach is to consider the archetypal five-atom cubic $AB$O$_3$ perovskite as a basic chemical unit, interleaving any two perovskites to form a layered bi-color superlattice structure.
[It is worth remembering that neither Glazer octahedral rotation patterns in bulk perovskites  nor cation layering of the  $A/A^\prime$ (or $B/B^\prime$) sites in such  bi-color superlattices are capable of individually producing noncentrosymmetric structures.]
We consider $(A B$O$_3)_1/(A^\prime B$O$_3)_1$ and $(AB$O$_3)_1/(AB^\prime $O$_3)_1$ superlattices, corresponding to the bulk-like compositions $(A,A^\prime)B_2$O$_6$ and $A_2(B,B^\prime)$O$_6$ (Sec.~SI) with layered cation-orderings (Fig.~\ref{fig:ingredients}),  as the simplest  structures that allow us to formulate materials selection rules in terms of the properties of the two {\it bulk} constituents: The chemical composition of the basic perovskite units considered, 
and the energetic or lattice dynamical  properties of the five-atom perovskite units.

%
{\sffamily Chemical Criterion:   
The synthetic perovskites require an $A/A^\prime$ layered cation ordering  for the octahedral rotations to induce ferroelectricity.}
The crystallographic rational for this criterion is as follows: In the $AB$O$_3$ perovskite structure both the $A$-site and the $B$-site positions have local inversion  symmetry (Fig.~\ref{fig:ingredients}a). Rotations can lift the inversion center on the $A$-site, but always preserve the centricity (inversion on the $B$-site) of the octahedra. $A/A^\prime$ ($B/B^\prime$) layered cation ordering removes the inversion center on the $B$-site ($A$-site) as shown in Fig.~\ref{fig:ingredients}b and c. Therefore the combination of $A/A^\prime$ ($B/B^\prime$) cation order and rotations allows for (forbids) hybrid improper ferroelectricity. 

This heuristic understanding is verified by our group theoretic methods. We find seven unique tilts patterns obtained from pairs of centric $B$O$_6$ octahedral rotations patterns in the paraelectric $(A,A^\prime)B_2$O$_6$ structure. Each combination leads to a $P(Q_1 Q_2)$ term in the free energy (Table S5) allowing for rotation-induced ferroelectricity---neither SOJT cations nor polar cation displacements are required. Note that the form of the trilinear term indicates that the combined rotation pattern is effectively polar; once the two rotations are present, there is only one direction for the induced polarization, i.e., turning on the two rotations is analogous to turning on an electric field.
In contrast, we find no trilinear invariants in the paraelectric $A_2(B,B^\prime)$O$_6$ structure and therefore hybrid improper ferroelectricity is forbidden.

\begin{figure}[t]
\centering
\includegraphics[width=0.495\textwidth]{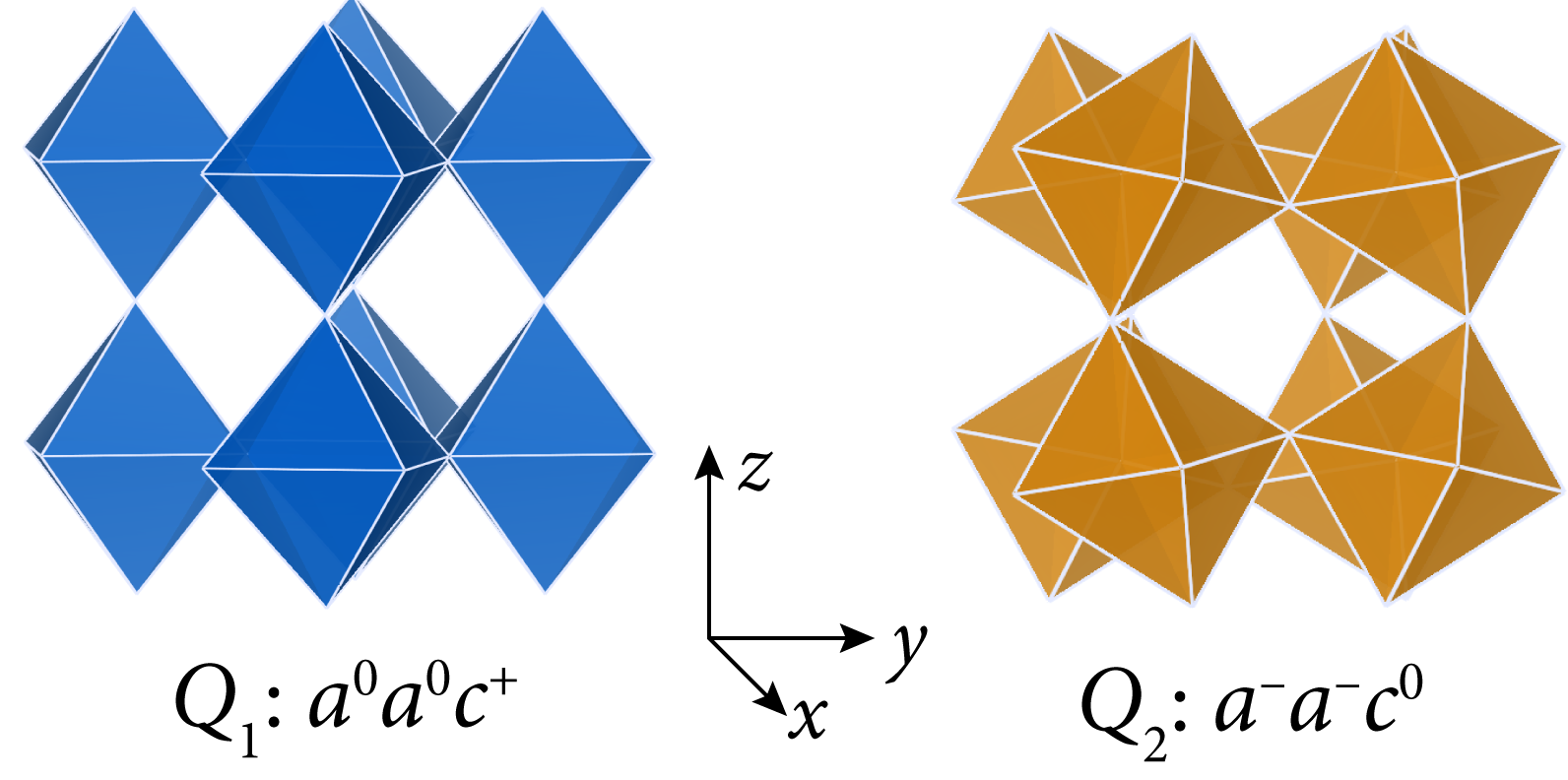}\vspace{-9pt} 
\caption{\label{fig:ortho_tilt}The primary rotational modes 
$Q_1$ ($a^0a^0c^+$) and $Q_2$ ($a^-a^-c^0$)  
targeted to combine in the $A/A^\prime$ superlattices to induce hybrid 
improper ferroelectricity. The combination of these rotations produces the 
common orthorhombic ($a^-a^-c^+$) tilt pattern found in perovskites.
}
\end{figure}
%
The chemical criterion is insufficient to guarantee that the octahedral rotations \textit{induce} a polarization, i.e., that $Q_1 Q_2 \rightarrow P$.
Rather, the form of the trilinear coupling, $P (Q_1  Q_2)$, compels only the mutual coexistence of such lattice distortions, which are commonly incompatible with one another, if any two are unstable.\footnote{
In many cases, 1/1 superlattices that satisfy this condition can be constructed from 
basic perovskite chemical units that are ferroelectric in the bulk, e.g.\ PbTiO$_3$:   $P$ then describes a conventional (soft transverse optical mode) ferroelectric instability thats coexists with the octahedral rotations, rather than emerging from them.
It is therefore equally possible that the polarization and a single octahedral rotation distortion are the primary instabilities that break inversion symmetry and induce a second rotational distortion,\cite{Perez-Mato/Parlinski:2011} i.e., $P  Q_2 \rightarrow Q_1$ (or $P  Q_1 \rightarrow Q_2$); this does not yield octahedral rotation-induced HIF.}

%
{\sffamily Energetic Criterion: 
The ground state structure of the $AB\textrm{O}_3$ \textit{and} the $A^\prime B\textrm{O}_3$ building blocks must each contain the corresponding $Q_1$ and $Q_2$ rotation patterns.}
This requirement provides an optimal condition for hybrid improper ferroelectricity: it ensures the two octahedral rotations in the synthetic superlattice dominate the energy landscape over other competing instabilities, indicating they drive the transition to the polar structure.
Said another way, the ground state structure of both bulk chemical units should possess the rotation pattern which results from the specific combination of octahedral rotations modes, $Q_1$ and $Q_2$, that are desired to be present in the 1/1 $AB$O$_3$/$A^\prime B$O$_3$ superlattice.

This criterion stems from the fact that the rotation patterns are fully coherent in the synthetic $A/A^\prime$ structures.
A rotational instability that appears in only one of the constituents, although likely capable of inducing the same rotation pattern in the second chemical unit,\cite{Rondinelli/Spaldin:2010b} would produce a tilt pattern with overall smaller  octahedral rotations and modulated rotation angle amplitudes.
Such compound tilt patterns are detrimental to cooperative ordering of local dipoles and could hinder octahedral rotation-induced ferroelectricity.\\

%
\noindent %
{\sffamily \bfseries First-principles guided materials 
design$\quad$}\\ \noindent
Here we apply our criteria in order to select bulk single phase perovskite oxides with suitable 
chemical compositions and rotational patterns for integration 
into the 1/1 ordered $AB$O$_3$/$A^\prime B$O$_3$  perovskites. 
The first design criterion is necessarily satisfied by restricting our investigation to this composition. 
Next we apply group-theoretic methods and consider all symmetry-adapted normal mode patterns (Table S5) that produce octahedral tilt patterns in the paraelectric ($AB$O$_3$)$_1$/($A^\prime B$O$_3$)$_1$ structure. This analysis reveals many different possible invariants (Eq.\ S1) that permit the octahedral rotations to induce ferroelectric polarizations through the trilinear $P Q_1  Q_2$ term in the free energy.
To narrow the search, we focus on the lattice modes $Q_1 =a^0a^0c^+$ and $Q_2=a^-a^-c^0$ (Fig.~\ref{fig:ortho_tilt}).
These rotational modes are prime distortions to target because 
they are ubiquitous in single phase perovskites; they produce orthorhombic $Pnma$ perovskites with 
the $a^-a^-c^+$ octahedral tilt pattern.

We choose to focus on five different gallate and aluminate perovskites: YAlO$_3$, LaAlO$_3$, BiAlO$_3$, LaGaO$_3$, and YGaO$_3$. We evaluate the lattice stability from calculations of the force constants throughout the Brillouin zone of these bulk compounds (Fig.~S2) and perform (zero kelvin) full structural relaxations in various isotropy subgroups corresponding to the condensation of one or more of the unstable modes at the high-symmetry points within the Brillouin zone.
In Table~\ref{tab:bulk_perovskites} we enumerate the most unstable phonon modes  at the high symmetry points in the tetragonal Brillouin zone of the non-polar ($P4/mmm$) bulk phases and identify the rotational pattern of the perovskite ground state.
Note, although some of these bulk compounds show polar $\Gamma$-point instabilities,\footnote{Although complicating the analysis, if a bulk material is a 
proper ferroelectric, the presence of a soft polar mode does not preclude the 
possibility of octahedral rotation-induced ferroelectricity.}  
none of them exhibit ferroelectric perovskite ground state structures in the 
bulk except for $R3c$ BiAlO$_3$.

The combination of our targeted octahedral rotation modes,  $Q_1=a^0a^0c^+$ ($M$-point) and $Q_2=a^-a^-c^0$ ($A$-point), determine the ground state tilt patterns of three of the five compounds: 
LaGaO$_3$, YGaO$_3$ and YAlO$_3$ all exhibit the orthorhombic $a^-a^-c^+$ rotation pattern at room temperature (Fig.~{\ref{fig:perovskite}c).
In contrast, LaAlO$_3$ and BiAlO$_3$ display the rhombohedral  $a^-a^-a^-$ tilt pattern.
\begingroup
\squeezetable
\begin{table}[t]
\begin{ruledtabular}
\centering
\caption{\label{tab:bulk_perovskites}Phonon modes of the single 
  phase paraelectric  aluminates and gallates used in 
  the selection of bulk perovskites to combine and realize 
  rotation-induced ferroelectricity.
	Calculated frequency and distortion-types  
	of the most unstable phonon modes for the 
	reference $P4/mmm$ bulk phases of the constituent superlattice 
	materials. Imaginary frequencies indicate energy lowering 
	instabilities.
	The high-symmetry $k$-points have the following wave vectors:
	$\Gamma=(0,0,0)$, $M=(\frac{1}{2},\frac{1}{2},0)$, 
	$R=(0,\frac{1}{2},\frac{1}{2})$, and $A=(\frac{1}{2},\frac{1}{2},\frac{1}{2})$.
	%
	}	
\begin{tabular}{clcl}%
Material	& & & \\ 
(bulk tilt system) & $\omega$ (cm$^{-1}$)	& $k$-point	& distortion-type \\
\hline
LaAlO$_3$			& 46.9$i$ 		& $M$ 		& $a^0a^0c^+$ \\
($a^-a^-a^-$)				& 135$i$		& $A$		& $a^0a^0c^-$ \\
				& 108$i$		& $A$		& $a^-a^-c^0$ \\
\hline
BiAlO$_3$		& 175$i$	& $\Gamma$& $z$-polarization\\
($a^-a^-a^-$)		& 158$i$	& $\Gamma$& $xy$-polarization \\
				& 234$i$ 		& $M$ 		&  $a^0a^0c^+$\\
				& 211$i$		& $R$		&  $a^+a^+c^0$ \\
				& 247$i$		& $A$		&  $a^0a^0c^-$\\
				& 232$i$		& $A$		&  $a^-a^-c^0$\\
\hline
YAlO$_3$		 &	 96$i$	& $\Gamma$& $z$-polarization	\\
($a^-a^-c^+$)		        & 248$i$ 		& $M$ 		&  $a^0a^0c^+$ \\
				& 229$i$		& $R$		&  $a^+a^+c^0$ \\
				& 291$i$		& $A$		&  $a^0a^0c^-$ \\
				& 282$i$		& $A$		&  $a^-a^-c^0$ \\
\hline
LaGaO$_3$		& 177$i$ 		& $M$ 		&  $a^0a^0c^+$ \\
($a^-a^-c^+$)			& 165$i$		& $R$		&  $a^+a^+c^0$ \\
				& 215$i$		& $A$		&  $a^0a^0c^-$ \\
				& 209$i$		& $A$		&  $a^-a^-c^0$ \\
\hline
YGaO$_3$	 		& 180$i$	& $\Gamma$& $z$-polarization	\\
($a^-a^-a^+$)\footnotemark[1]			& 144$i$	& $\Gamma$& $xy$-polarization	\\
				& 297$i$ 		& $M$ 		&  $a^0a^0c^+$ \\
				& 286$i$		& $R$		&  $a^+a^+c^0$ \\
				& 327$i$		& $A$		&  $a^0a^0c^-$\\
				& 322$i$		& $A$		&  $a^-a^-c^0$ \\
\end{tabular}
\end{ruledtabular}
\footnotetext[1]{When confined to a perovskite manifold of structures.}
\end{table}
\endgroup
We consider four representative 1/1 superlattice test cases based on these bulk candidate perovskites (Table~\ref{tab:results}):
\begin{itemize}
\item[(1)] YAlO$_3$/YGaO$_3$.---Both YAlO$_3$ and YGaO$_3$ exhibit strong $a^0a^0c^+$ and $a^-a^-c^0$  instabilities, i.e., the  targeted orthorhombic rotation pattern, but the $B/B^\prime$ ordering does not satisfy the chemical criterion, indicating that a trilinear coupling is symmetry forbidden. 
\item[(2)] LaGaO$_3$/YGaO$_3$.---To mitigate this issue, we substitute YAlO$_3$ with LaGaO$_3$ to recover the trilinear coupling enabled by the chemical criterion. Since we find strong  $a^0a^0c^+$ and $a^-a^-c^0$ rotations present in both compounds (Table~\ref{tab:bulk_perovskites}), we anticipate that the energetic criterion will be \textit{strongly} satisfied---rotation-driven ferroelectricity should be active in the LaGaO$_3$/YGaO$_3$ superlattice.
\item[(3)] LaAlO$_3$/YAlO$_3$.---Starting again from (1), we will satisfy the  chemical criterion by substitution of YGaO$_3$ with LaAlO$_3$. In contrast to case $(2)$, the energetic requirement in this all-aluminate superlattice is  \textit{weakly} satisfied because LaAlO$_3$ displays the rhombohedral tilt pattern in its ground state rather than the targeted orthorhombic one; this is reflected in the substantially harder $a^0a^0c^+$ and $a^-a^-c^0$ mode frequencies.
\item[(4)] LaAlO$_3$/BiAlO$_3$.---Finally, rotation-induced ferroelectricity should be deactivated by substituting YAlO$_3$ with BiAlO$_3$ because neither compound possesses the orthorhombic $Pnma$ tilt pattern---a direct violation of the energetic criterion despite satisfying the chemical restriction. Proper ferroelectricity
could emerge, however, from the strong polar instability
present in BiAlO$_3$.
\end{itemize}
\noindent
{\sffamily \bfseries Results$\quad$}\\ 
We find that all superlattice compositions with $A/A^\prime$ cation ordering have 
polar crystal structures  consisting of cation displacements 
and octahedral rotations (Fig.~S3).
The calculated ferroelectric polarizations shown in  Table~\ref{tab:results} 
are comparable to conventional ferroelectrics, e.g.\ the prototypical 
perovskite BaTiO$_3$ exhibits a  $\sim$33~$\mu$C~cm$^{-2}$ polarization in is ground state structure.\cite{Zhong/Vanderbilt/Rabe:1994}
Consistent with our materials guidelines, the $B/B^\prime$ cation ordered superlattice is centrosymmetric.
For the polar structures, however, the origin for ferroelectricity remains to be identified. 
\emph{The necessary condition to realize rotation-induced ferroelectricity is an interaction between different rotation distortions that produces a net energy gain to the polar ground state from their mutual coexistence.} We determine this interaction by analyzing the energy surface around the paraelectric parent structure in terms of the unstable lattice modes that produce the octahedral rotations present in the ground state structures. We show from this analysis that complete control over hybrid improper ferroelectricity is gained through application of our simple design criteria.
\begingroup
\squeezetable
\begin{table}[b]
\begin{ruledtabular}
\centering
\caption{\label{tab:results} Conditions for hybrid improper ferroelectricity (HIF) in 1/1 
perovskite superlattices.  
  For the octahedral rotations to induce a ferroelectric polarization in the 
  cation ordered superlattices both the chemical ($A/A^\prime$ ordering) and energetic 
  (orthorhombic $a^-a^-c^+$ tilt pattern) criteria   must be satisfied.} 
\begin{tabular}{rlcccc}\vspace{-0.3em}
			&																							& $A/A^\prime$ &	$a^-a^-c^+$ & polarization & \\
			&	&	ordering?	& tilt pattern? 	& ($\mu$C~cm$^{-2}$) & HIF? \\
\hline
($1$) & $\textrm{YAlO}_3/\textrm{YGaO}_3$ 	& no	& no					& 0	& no	\\
($2$)& $\textrm{LaGaO}_3/\textrm{YGaO}_3$ 	& yes	& yes, strong	& 11.8 & yes	\\
($3$)& $\textrm{LaAlO}_3/\textrm{YAlO}_3$	& yes	& yes, weak		& 9.94 & yes	\\
($4$)& $\textrm{LaAlO}_3/\textrm{BiAlO}_3$&	yes	&	no					& 9.62 & no	\\
\end{tabular}
\end{ruledtabular}
\end{table}
\endgroup

\begin{figure*}
\includegraphics[width=0.99\textwidth]{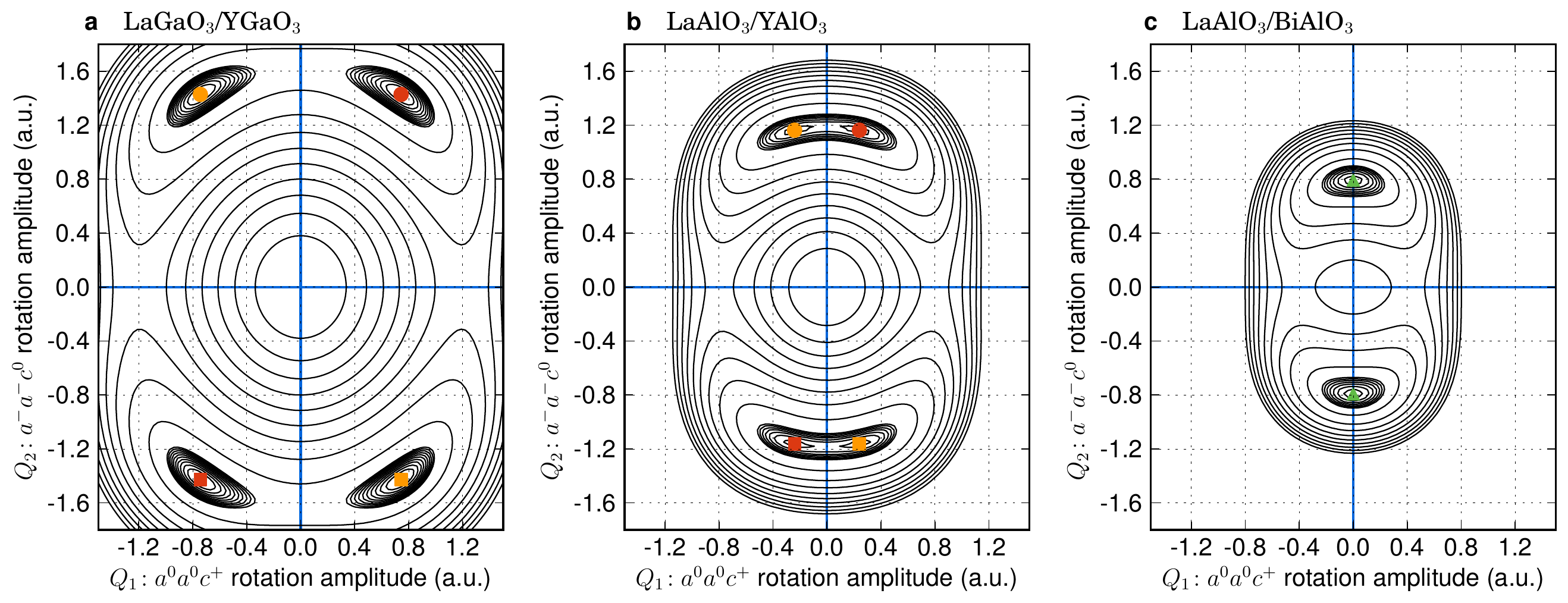}\vspace{-2pt}
\centering
\caption{\label{fig:2dplots}%
	Energetic signatures for rotation-induced ferroelectricity.
	The calculated zero kelvin two-dimensional energy surface contours for 
	each cation-ordered perovskite with respect to the primary 
	$Q_1\!: a^0a^0c^+$ and $Q_2\!: a^-a^-c^0$ rotations  
	centric octahedral rotation modes  
	present in the polar ground state structures. 	
	Filled symbols indicate the positions of energy minima in the potential 
	energy landscape.
	Equivalently colored symbols represent structures with 
	identical ferroelectric polarization directions.
	Circles and squares differentiate anti-phase (domain) 
	structures, which have identical polarization directions, but possess 
	different non-polar structural distortion directions.	
	In (a) and (b) the two $a^0a^0c^+$ and $a^-a^-c^0$ rotations 
	combine to produce 
	the orthorhombic $a^-a^-c^+$ tilt pattern: 
	(a) The strong rotational modes in the LaGaO$_3$/YGaO$_3$ superlattice 
	produce four polar crystal structures.
	Each $Pmc2_1$ structure is symmetry related to the others 
	as either ferroelectric twin structures, which differ in 
	their polarization direction ([110]- or [$\bar{1}\bar{1}0$]-type), or as 
	antiphase domains, which differ in the relative sign of the GaO$_6$ rotations while 
	the polarization direction remains fixed.	
	(b) Despite the $a^0a^0c^+$ rotation being substantially weaker than the 
	$a^-a^-c^0$ tilt in the LaAlO$_3$/YAlO$_3$ structure, 
	the combination of two non-polar octahedral rotations 	produce four symmetry related polar $Pmc2_1$ structures. The energy barrier 
	separating the ferroelectric twins, however, is substantially reduced. 
	(c) In the cation ordered LaAlO$_3$/BiAlO$_3$ superlattice this energy barrier 
	collapses to zero: the $a^0a^0c^+$ and $a^-a^-c^0$ strongly compete
	 with each other 
	to produce two symmetry equivalent non-polar (green 
	triangles) $Pmma$ structures, indicating the calculated ferroelectric 
	polarization is not induced by the octahedral rotations.
}
\end{figure*}

{$(1)~\textrm{YAlO}_3/\textrm{YGaO}_3$}.---%
The centrosymmetric ground state structure exhibits anti-parallel cation displacements and 
the centric $a^-a^-c^-$ tilt pattern. Here, adjacent octahedra rotate out-of-phase in all directions and the $xy$-rotation angle magnitude is modulated from Ga-layer to Al-layer along the $z$-direction (Fig.\ S4).
This octahedral motif results from a combination of the $a^0a^0c^-$ and $a^-a^-c^0$  unstable zone-boundary instabilities of the paraelectric superlattice. 
However, the combination of  $a^0a^0c^-$ 
(or the symmetry equivalent but energetically unique $a^0a^0c^+$ rotation) with the $a^-a^-c^0$ mode in the $B/B^\prime$ superlattices produces only \textit{centrosymmetric} structures (Fig.~S5) and therefore prohibits the octahedral rotations from inducing ferroelectricity. 
Consistent with this result, we find that these rotations also produce an identical non-polar 
structure in the Bi substituted analogue, BiAlO$_3$/BiGaO$_3$ (Fig.~S6).
Our first-principles predicted centrosymmetric $P2_1/c$ structures confirm  
the symmetry-derived chemical criterion: Compositions with $B/B^\prime$ cation ordering 
prohibit the centric octahedra from producing a ferroelectric polarization.
{$(2)~\textrm{LaGaO}_3/\textrm{YGaO}_3$}.---%
The ground state gallate superlattice exhibits the targeted orthorhombic 
$a^-a^-c^+$ octahedral tilt pattern  and possess polar, rather than fully compensated antiparallel, cation displacements.
Adjacent GaO$_6$ octahedra rotate coherently in-phase (with the same magnitude) 
about the axis perpendicular to the La/Y ordering and out-of-phase in the two orthogonal directions (Fig.~\ref{fig:ortho_tilt}), which result from the combination of the highly unstable 
$a^0a^0c^+$ and $a^-a^-c^0$  rotations present in the 
paraelectric LaGaO$_3$ and YGaO$_3$ phases (Table~\ref{tab:bulk_perovskites}) that 
are each independently energy lowering.
Rotation-induced ferroelectricity requires the two octahedral rotation modes present in the polar structure to dominate the energy landscape. 
To explore this, we map out the two-dimensional energy surface contours for the 
LaGaO$_3$/YGaO$_3$ superlattice in terms of these modes (Figure~\ref{fig:2dplots}a):
We find four symmetry  equivalent energy  minima (denoted by filled symbols)  with noncentrosymmetric $Pmc2_1$ crystal structures, indicating that the $a^0a^0c^+$ and $a^-a^-c^0$ rotations combine to produce the ferroelectric structure in the absence of cation displacements. If the two modes were strongly in competition with each other, producing an energy penalty by their mutual coexistence in the system, then only two minima would result.
The location of these minima deep inside the quadrants and the large energy barrier 
($\sim$39~meV/f.u.) separating the ferroelectric twins (at $Q_1 >0~\textrm{and}~<0$) 
reflects the strong favorable coupling between the largely unstable rotational modes possessed by both  bulk constituents. 
This barrier is further enhanced by nearly 3X by substitution of Y with Bi (Fig.~S7).
For these reasons, we anticipate rotation-induced ferroelectricity in this  
composition to be robust to thermal effects at room temperature.
To verify that the $a^-a^-c^+$ octahedral tilt pattern induces the electric polarization (and the subsequent cation displacements), we remove all polar distortions from the $Pmc2_1$ ground state structure  and calculate the electronic-only contribution to the total polarization using the Berry phase method with only the octahedral rotations frozen in.\cite{King-Smith/Vanderbilt:1993} 
Any electric polarization now must originate solely from 
the change in charge density induced by the cooperative and non polar oxygen displacements (octahedral rotations).
Here, we find the rotations 
induce a sizable electronic-only polarization of 0.65~$\mu$C~cm$^{-2}$.
The polarized charge density acts as an effective 
electric field, driving the ions to 
displace in a polar fashion---this is the essence of hybrid 
improper ferroelectricity. 
These induced cation displacements contribute 
to the total electric polarization in the fully relaxed $Pmc2_1$ structure. 
As a result, the polarization $P$ is 
intimately linked to the $a^-a^-c^+$ tilt pattern in LaGaO$_3$/YGaO$_3$.

{$(3)~\textrm{LaAlO}_3/\textrm{YAlO}_3$}.---%
The ground state polar structure of the all-aluminate composition is 
identical to the previous all-gallate superlattice: it  
exhibits the targeted $a^-a^-c^+$ tilt pattern and polar cation displacements (space group 
$Pmc2_1$). 
As before, we find four symmetry  equivalent energy  minima resulting from the combination and cooperation of the 
$a^0a^0c^+$ and $a^-a^-c^0$ rotations {\it cooperate} with one another (Fig.~\ref{fig:2dplots}b).
After full structural relaxation of this rotationally-only distorted structure 
(corresponding to the minima in Fig.~\ref{fig:2dplots}b), we find polar cation displacements 
occurring in response to the  $a^-a^-c^+$ tilt pattern through the symmetry allowed trilinear $P_z  (Q_1 Q_2)$	coupling term absent in single phase non-polar perovskites.
Although the net interaction between the two rotation modes lower the energy of the LaAlO$_3$/YAlO$_3$ system, $Q_2$  is stronger than $Q_1$. This is evident from inspection of the unstable phonon modes for the bulk constituents (Table~\ref{tab:bulk_perovskites}) and is also discernible in 
Fig.~\ref{fig:2dplots}b, where the energy minima are located close to the $Q_1=0$  boundary.
This weak susceptibility to the $a^0a^0c^+$ tilt system 
in paraelectric LaAlO$_3$, leads to 
a small ($\sim$1~meV/f.u.)  zero-kelvin energy barrier separating the two ferreoelectric twins.
Thus the energetic balance between the two rotation 
modes and degree to which the energetic criterion is satisfied is equally as important as  
maintaining an $A/A^\prime$ cation ordered superlattice (chemical criterion).
It determines the stability of octahedral rotation-induced HIF: 
Because the   $a^0a^0c^+$ rotation instability  in LaAlO$_3$ is weak, its presence in the ground state crystal structure is highly susceptible to finite temperature effects.
If the four polar minima are thermally suppressed then two non-polar $Pmma$ crystal structures would result, subverting the energetic criterion and nullifying the rotation-induced ferroelectricity mechanism. 
%

{$(4)~\textrm{LaAlO}_3/\textrm{BiAlO}_3$}.---%
The ground state polar structure exhibits the targeted $a^-a^-c^+$ tilt pattern; however, 
Figure~\ref{fig:2dplots}c illustrates that  in the absence of polar ionic displacements, 
the energy interaction between the two 
$a^0a^0c^+$ and $a^-a^-c^0$ rotations disfavors their mutual coexistence. 
Only two symmetry equivalent energy minima 
with  centrosymmetric $Pmma$ structures occur.
In this extreme case, the energy cost of the coupling between rotations exceeds the gain from the individual modes. 
In other words, while the rotations $Q_1$ and $Q_2$ are of suitable symmetry to induce an electric polarization and are individually unstable (Table S4), our density functional calculations reveal that \textit{one} mode dominates and suppresses the other. 
This happens in part due to the weak $Q_1$ instability  in LaAlO$_3$ and  strong anharmonic coupling between  $Q_1$ and $Q_2$ in bulk BiAlO$_3$. Both of these features are reflected in the fact that neither lanthanum aluminate nor bismuth aluminate display the desired $a^-a^-c^+$ tilt system in their ground states.  The ferroelectric polarization that results in this system is due to a conventional soft polar mode; rotation-induced ferroelctricity is absent in the $\textrm{LaAlO}_3/\textrm{BiAlO}_3$ system.\\

%
\noindent %
{\sffamily \bfseries Outlook and summary$\quad$}\\ \noindent
Finally, we describe important applications exploiting the trilinear coupling 
that enables centric octahedral rotation-induced ferroelectricity.
%
This mechanism supports electric polarizations in more diverse chemistries.
Unlike the conventional SOJT-mechanism that lifts inversion symmetry and produces electric polarizations through cooperative displacements of cations with $d^0$ electronic configurations (group 4, 5 and 6 transition metal ions) or cations with stereochemical active $ns^2$ lone pair  electrons, no such restriction is imposed on the cations' valence in this form of hybrid-improper ferroelectricity. 
Cations with strong magnetic interactions, open $d$- or $f$-shell configurations,   which are  incompatible  with conventional ferroelectricity,\cite{Hill:2000} are able to both fully coexist and couple to sizable electric polarizations:  
Centric octahedral rotation-induced ferroelectricity provides a plausible route to achieving robust magnetoelectric multiferroics. 

It is interesting to conjecture that there 
are viable chemistries providing rotation-induced ferroelectrics with anomalously large {\it electronic-only} contributions to the total polarizations. In cases where the polarization contribution (and distortions) due to ionic displacements is also small, reversal of the electric polarization would require minimal ionic motion. These materials would satisfy many of the  robust cycling and ultra fast (femtosecond timescale) dynamical switching requirements needed for next-generation electronics.

We have used first-principles density functional calculations combined with group theoretical studies to enumerate the crystal--chemistry design criteria required for the centric octahedral rotations pervasive in perovskite oxides to induce ferroelectric polarizations. 
We showed that interleaving two bulk perovskites to form an ordered and layered arrangement of $A/A^\prime$ cations [chemical composition $(A,A^\prime)B_2$O$_6$] produces a new trilinear free energy term coupling three lattice modes: 
two  octahedral rotations that dominate the energy landscape, forming  the orthorhombic $a^-a^-c^+$ perovskite tilt system, and an electric polarization. 
For ordered perovskites where these conditions are satisfied, the trilinear term induces an electric polarization  and hybrid improper ferroelectricity results. 
Because most octahedral rotations in perovskites freeze-in at elevated temperatures ($>$300~K), we argue that the trilinear coupling provides a robust route to realize synthetic ferroelectrics from constituents which are not ferroelectric in the bulk at room temperature.
By leveraging systematic symmetry analysis with first-principles density functional calculations, we illustrated an {\it a priori} materials selection strategy for designing synthetic ferroelectric crystals only from knowledge of the lattice dynamics of the constituent materials.\\

\noindent %
{\sffamily \bfseries Methods$\quad$}\\ \noindent
Our zero kelvin density functional calculations are performed within 
the local density approximation (LDA) using the 
Vienna {\it ab initio} Simulation Package ({\sc vasp})
\cite{Kresse/Furthmuller:1996a,Kresse/Joubert:1999} and 
the projector augmented wave (PAW) method \cite{Blochl:1994}
with the following valence electron configurations:
$5s^25p^65d^16s^2$ (La), 
$4s^24p^65s^24d^1$ (Y), 
$3d^{10}4s^2 4p^1$ (Ga), 
$3s^2 3p^1$ (Al) 
and $2s^22p^4$ (O).
For the reference $P4/mmm$  bulk perovskite structures  
we use a $5\times5\times5$ Monkhorst-Pack 
$k$-point mesh~\cite{Monkhorst/Pack:1976} and a 500~eV plane wave cutoff.
We then volume-optimize each structure within $P4/mmm$ symmetry.
For our superlattice calculations we use a $5\times5\times3$ Monkhorst-Pack 
$k$-point mesh and a 550~eV plane wave cutoff.
We relax the ionic coordinates until the Hellmann-Feynman forces on the 
atoms are less than 0.1 meV~\AA$^{-1}$.
We obtain the phonon frequencies at high symmetry Brillouin zone points 
(Table \ref{tab:bulk_perovskites}) by calculating total energies with respect to 
atomic displacements about the reference $P4/mmm$ structure in a 2$\times$2$\times$2 
supercell.
For these calculations we use a larger, 650~eV, planewave cutoff.
In this frozen-phonon method, a series of small (symmetry inequivalent) atomic
displacements are imposed along different Cartesian directions.
We calculate the dynamical matrix from the Hellman-Feynman forces and 
diagonalize the dynamical matrix to obtain the atomic displacement patterns
(eigenvectors) and phonon mode frequencies (eigenvalues).
We obtain the ground state atomic structures for all cation ordered perovskite 
compositions by 
systematically ``freezing-in'' linear combinations of the unstable phonon modes 
(Table~\ref{tab:bulk_perovskites}) into the paraelectric cation-ordered 
perovskite reference structures (Sec.~SII.B) and then performing full structural relaxations from first-principles.\\

\noindent %
{\sffamily \bfseries Acknowledgments$\quad$}\\ \noindent
The authors thank N.A.\ Benedek, E.\ Bousquet, A.~Cano, K.R.\ Poeppelmeier, M.\ Stengel and K.~Rabe for useful discussions 
and support from  U.S.\ DOE, Office of Science, under Contract No.\ 
DE-AC02-06CH11357 (JMR) and the DOE-BES under Grant No. DE-SCOO02334 (CJF).
Computational support was 
provided by the DOE-BES, through the high-performance computing 
facilities at the Center for Nanoscale Materials and the LCRC operated 
FUSION compute cluster at Argonne National Laboratory. This work 
was initiated during a stay at the AQUIFER Program of the IMI and 
NSF under Award no. DMR-0843934, managed by the International 
Center for Materials Research, University of California, Santa Barbara, 
USA.

\bibliography{rondo}

\end{document}